\documentclass[twocolumn]{revtex4}
\usepackage [dvips]{graphicx}
\def\D#1#2{\frac{\partial #1}{\partial #2}}

\def\vec#1{\mathbf{#1}}

\begin{document}
\title{Space Dynamics in Global Time as an Effective Alternative
to General Relativity}%
\author{Dmitry E. Burlankov}%
\email{bur@phys.unn.runnet.ru}
\affiliation{Physics Department, University of Nizhny Novgorod, Russia.}
%Authors' institution and/or address}
\begin{abstract}
The fundamental physical object of the {\it Global Time Theory} is a
three-dimensional curved space dynamically developing in global time.
The equations of its dynamics are derived from the Lagrangian, and the Hamiltonian of gravitation
turns out to be nonzero.
The General Relativity solutions are shown to be a subset of the GTT solutions with zero energy
density.
In Global time Theory, the quantum theory of gravitation can be built on the basis of the Schr\"odinger
equation, as for other fields. The quantum model of the Big Bang is presented in some detailes.
\end{abstract}

%\pacs{04, 04.60.-m, 98.80.Hw}
\maketitle

\tableofcontents

\section{Dynamic geometry}

{\bf General relativity} (GR) is built on the Riemannian geometry of space-time \cite{Wheeler,LL}.
Solutions of its basic equations, the Einstein equations, determine full four-dimensional space-time,
past, present and future simultaniously. In contrast, {\bf The Global Time Theory} (GTT) considers
a three-dimensional configurational space as a fundamental physical object that develops dynamically
in {\it global time} which is common for all points of the space.

From times of Ancient Greece, geometry was build on the basis of observations on
positions and motion of objects with respect to each other, without any connection to time.
The geometry of Euclid and Lobachevsky, and the geometries of Gauss and
Rieman do not contain the concept of time.
These geometries are intended for undisturbed or slowly moving observers contemplating
placed in front of them blueprints or still objects.

The Newton mechanics has introduced to mathematical descriptions of nature the notion of
{\it time}. However, Newton assumed space to be Euclidean what was natural for his time.
Therefore, introduced by him moving systems also were Euclidean spaces moving in
 {\it absolute space} (since the Euclidean space admits, from
the point of view of modern mathematics, {\it motion}).

The very first description of dynamics with inhomogeneous  velocity  fields
belongs to Euler.
In the process of formulating his famous hydrodynamical equations, Euler applied in
a local frame connected to the moving fluid the second Newton's law:
$$\D{\vec{v}}{t}=-\frac{1}{\rho}\,\vec{\nabla}p,$$
Then he transferred the time derivative from a system in which the fluid elementary volume is at rest
($\vec{v} =0$), to the laboratory system replacing the time derivative by the so called
{\it substantial derivative}
\begin{equation} \label{Eiler}
 \D{}{t}\to\D{}{t}+(\vec{v}(\vec{r},t)\cdot\vec{\nabla}).
\end{equation}
However, such transformation is valid for a scalar quantity and (by a coincidence)
 for field of velocities.

The anzats for transformation of tensors of any rank due the time shift
was prepared by Sophus Lie
when he introduced for an arbitrary infinitesimal transformations of coordinates
 $\bar{x} ^i=x^i-\xi^i (x)$ so called
{\it a Lie-variation} of a tensor field (for example, for the covariant vector
field $A_i$):
\begin{equation} \label{Lie}
  \delta_\xi\,A_i=\xi^s\,A_i,_s+\xi^s,_i\,A_s.
\end{equation}

In {\bf Dynamic geometry} the space is represented by a set of all its points.
The frame of references in which
coordinates of the space points ($ \bar{x} ^i $) do not change with time, we shall term
{\it as an absolute inertial frame}. Then, in some other frame of references
where point coordinates are connected to their coordinates in an absolute
inertial frame by a time-dependent transformation $ x^i=f^i (\bar{x}, t)$, there
exists {\it a field of absolute velocities}
\begin{equation} \label{Vabs}
V^i =\D{x^i}{t},
\end{equation}
absent in an absolute inertial frame (what is the distinct feature of
an absolute inertial frame).
An infinitesimal increment of time $dt $ generates an infinitesimal variation
of coordinates $ \xi^i (x, t) =V^i (x, t) \, dt $. The latter transforms the time derivative from a tensor field,
taking into account the Lie variation (\ref{Lie}),
from the non-inertial frame to the inertial one, forming the {\it covariant derivative
of the tensor field with respect to
Time} (\cite{book}).
The derivative contains $r+2 $ elements, where $r $ is a rank of the tensor. In particular,
for a scalar field ($r=0$)
$$D_t\,f(x,t)=\D{f}{t}+V^i\,\partial_i\,f,$$
and for a covariant field of vectors $A_i(x,t)$
$$D_t\,A_i=\D{A_i}{t}+V^s\,A_i,_s+V^s,_i\,A_s.$$

Especially important in the dynamics of space is the time covariant derivative
from a metric tensor
\begin{equation} \label{gcov} D_t\,\gamma_{ij}=\D{\gamma_{ij}}{t}+V_{i;j}+V_{j;i}.
\end{equation}

\section{The Global Time Theory}

The Global Time Theory originates from the following physical concepts of space
and time:
\begin{description}
\item [Space] is the material carrier of geometrical properties. It is three-dimensional.
\item [Global time] is a proper time of the space, uniform for all of its points.
It flows everywhere and always equally uniformly, it is itself the measure of the
uniformity.
\end{description}

The space is {\it the carrier of geometrical properties} because geometrical
properties are determined by a metric tensor, six components of which are major
field variables of the space.

Bodies move in space, fields dynamics (for example, electromagnetic field)
develop in space. For each moving point {\it the absolute velocity}
relative to the space is defined.

There exists an absolute motion with respect to the space, or from different perspective,
there exists a velocity field in some frame of references.
Thus, the dynamics of space is described by six components of the metric tensor
field $ \gamma _{ij} (x, t) $, determining its geometrical properties in each given moment of time,
 and by three components of absolute velocities $V^i (x, t) $,
determining how each space point moves in each moment of time with respect to the
chosen frame of references.

The space is the {\it material}  carrier of geometrical properties, because
 equations of the dynamics of the metric tensor and the velocity field are obtained
from the Lagrangian equations and define, along with other fields (for example,
electromagnetic), energy of the system.

The metric dynamics and velocity field equations are derived from a
variation principle, where the gravitational action is presented as a difference
between
kinetic (quadratic with respect to velocities of the metric deformation) and potential energy
of the space (proportional to the scalar curvature of space):
\begin{equation}
 S=\frac{c^4}{16\,\pi\,k}\int(\mu^i_j\,\mu^j_i-(\mu^j_j)^2+R)\sqrt{\gamma}\,d_3\,x\,dt
 + S_m.
\label{GrAction}
\end{equation}
Here $S_m $ is the action of all other matter, and $\mu _{ij}$ is a tensor
of space deformation velocities consisting of time covariant derivatives
of the metric tensor (\ref{gcov}):
\begin{equation}
\mu _{ij} = \frac{1}{2 \, c}\,D_t\,\gamma _{ij} =
\frac{1}{2 \, c} (\dot{\gamma} _{ij} + V _{i; j} +V _{j; i}). \label{muij}
\end{equation}
Note that the action contains the field of absolute velocities  $V^i $ but only
through the tensor above.

Variation of the action with respect to the metric tensor $\gamma _{ij}$ leads to
six dynamic equations. Also variation with respect to the absolute velocity field
generates three {\it to the equations of connectivity}. Thus the set of equations of
the space dynamics consists of {\it nine} partial differential equations
of the second order.

From the action the energy density and the Hamiltonian
$H_\gamma=\int\rho\,\sqrt{\gamma}\,d_3x$ are determined as usually in field theory:
\begin{equation}\label{HamGr}
\rho=\frac{c^4}{16\,\pi\,k}\,(\mu^i_j\,\mu^j_i-(\mu^j_j)^2-R).
\end{equation}

An important feature of this Hamiltonian is its indeterminate sign .

\subsection{The proper time of the moving observer}
In GR, as Einstein constantly emphasized,
the special relativity theory is valid in local:
in small, the space and time are described by the Minkowski
metric (tangential space-time).

The global construction of GTT does not superimpose any restrictions on local properties.
The special relativity theory, as local structure of space-time, can be also naturally
incorporated in GTT. For any moving observer its proper time is defined via a global time
and velocity with respect to the space $v^i =\dot{x} ^i-V^i $:
\begin{equation} \label{tself}
d\tau=dt\,\sqrt{1-\frac{1}{c^2}\,\gamma_{ij}\,v^i\,v^j}.
\end{equation}

\section{Similarities between GTT and GR}

\subsection{ADM-representation}

Similarity and differences between GTT and GR can be the best traced in so called
ADM-representation of GR. Arnovitt, Deser and Misner \cite{ADM} have expressed ten
components of the four-dimensional metric tensor through six components of the metric
tensor of the three-dimensional space $ \gamma _{ij} $, the three-dimensional vector
$V^i$ (in notations of GTT), and the function{\it of a course of time} $f(x, t)$:
\begin{equation}
g _{00} =f^2-\gamma_{ij} V^iV^j;\quad g _{0i}=\gamma_{ij} V^j; \quad
g _{ij} = -\gamma _{ij}. \label{mtrADM}
\end{equation}

Components of an inverse metric tensor are respectively
\begin{equation}
g ^{00} = \frac{1}{f^2};\quad g ^{0i}=\frac{V^i}{f^2};
\quad g ^{ij} = \frac{V^iV^j}{f^2} -\gamma ^{ij}.
\label{obrmtrADM}
\end{equation}
Ten Einstein equations are obtained then as variation equations for
ten components of the metric tensor $g _{\alpha\beta} $ of the Hilbert action
\begin{equation} \label{Hilb}
S_G =\frac{c^4}{16\pi \, k} \, \int R \,\sqrt{g} \, d_4x,
\end{equation}
where $R $ is a scalar curvature of the {\it four-dimensional} space-time.

In GTT, the component $g ^{00} =1$ always and everywhere. This is a function
defining the pace of a global time. With this condition, the Hilbert action (\ref{Hilb})
becomes the action of a configurational space (\ref{GrAction}).
The component $g ^{00} $ can not be varied,
and the function which is multiplied by this variation can be arbitrary.
It is the function which is an energy density (\ref{HamGr}).

The variation of all ten components of the space-time metric leads to
the basic difference between expressions in GR and GTT:
the variation with respect to $f$ produces an extra
(comparison with nine equations of GTT)
 equation
\begin{equation} \label{zeroH}
H=0.
\end{equation}
The density of a complete Hamiltonian (including both space and substance) is equal then
to zero, and consequently the Hamiltonian itself is equal to zero in GR.

Solutions of GR therefore define a subset of all GTT solutions with
an energy density equal to zero everywhere. It is the tenth equation, in addition to six dynamic
equations and three connectivity equations, which cuts out the GR sector from GTT.

\subsection{Reduction of GR to a global time}

Cosmological solutions of GR always originate from the metric of the form:
$$ds^2=c^2\,dt^2-\gamma_{ij}(x,t)\,dx^i\,dx^j.$$
Here the component of the metric $g ^{00} =1 $ what means that the time is global, and
$g ^{0i} =0$ what means that the system is globally inertial.

The other solutions of GR (geodesically complete) also can be reduced
to the global time.
If there is a four-dimensional metric $g _{\alpha\beta} $ in arbitrary coordinates
$x ^\alpha $, in order to reduce it to a global time one needs to transform coordinates
(to pick only new time coordinate $ \tau=c \, t $, more precisely) so that the
requirement $g ^{00} =1 $ is satisfied. Following tensor transformation rules,
\begin{equation} \label{HJGTT}
\bar{g}^{00}=g^{\alpha\beta}\D{\tau}{x^\alpha}\D{\tau}{x^\beta}=1.
\end{equation}
This differential equation for $ \tau $ turns out to be the Hamilton-Jacobi
equation for trajectories of freely falling mass points (laboratories) for which a
common natural time is $t $.
Thus, {\it a principle of equivalence}, tied an inertial system
to the freely falling laboratory, exists in global time. However,
in contrast to the Einstein's elevator, there exists a multitude of such laboratories
and the time in them is synchronized.
Thus the principle of equivalence becomes global. But the three-dimensional manifold
formed by these points-laboratories,is not any more a Euclidean space.

In GTT, the theorem is proved that any static spherically symmetric solution
with any kind of matter has an energy equal to zero \cite{book}.
Therefore all such solutions in GTT and GR can be reduced to each other.
For example, the Shwarzshield  solution in global time can be presented as
\begin{equation} \label{gShGlob} ds^2=\left(1-\frac{2\,M}{r}\right)\,dt^2+2\sqrt{\frac{2\,M}{r}}\:dt\:dr\end{equation}
$$ (dr^2+r^2(d\vartheta^2+\sin^2\vartheta\:d\varphi^2)).$$

This expression was obtained from the Shwarzshield metric by Painleve in 1921
\cite{Painleve}.
He carried out in the Shwarzshield metric various transformations
$\tilde{t}=t+\Phi(r)$ and showed, that the crossections $t=const $ are different
at different choices of the function $\Phi(r)$.
In particular, he has found a function, at which the spatial crossection is an
Euclidean space. This solution can be obtained as a solution of the GTT equations
\cite{book}.

Solving the equation (\ref{HJGTT}), it is not difficult to reduce to global time
($g ^{00} =1 $) and other GR solutions, for example, the Nordstr\"om or
Kerr metric.

\section{Differences between GTT and GR}

As was shown above, solutions of GR form a subset of solutions of GTT with an energy density
equal to zero. The removal of this restriction leads not only to the appearance
of new solutions, but also to the removal of many problems of GR, such as the problem
of initial conditions, the problem of a critical density, geodesic completeness,
and return to the Shr\"oedinger theory in quantum area.

\subsection{Cosmological models}

For Friedman's model of space, the three-dimensional sphere of variable radius $r(t)$,
the energy (\ref{HamGr}) in a Planck units system is negative and is given by:
$$ E =-3 \, r \,\dot{r} ^2-3 \, r. $$

Since due to dynamical equations the energy is conserved, it is a differential
equation of the first order,the solution for which is a Friedman's cycloid:
$$\quad r =\frac{r_m}{2} (1-\cos\chi); \quad
t = \frac{r_m}{2} (\chi-\sin\chi). $$
The key difference here from the Friedman solution is only physical: in this solution
the matter density is absent. Presence of a dust-like matter changes only
the constant $E$. Thus in GTT, {\it a problem of a critical density}
in a cosmology is absent: the World can be closed or open independently from
the matter density.

\subsection{Space dynamics}

The removal of the GR restriction about the zero energy density
leads to solutions important for space dynamics: to the field of space vortexes
\cite{grVort}. These solutions have surprisingly simple mathematical properties
(weak superposition principle) and huge energies. Instead of hypothetical
"dark matter", "dark energy",  and "huge black holes ", significant energetic role
in space dynamics plays, from the GTT point of view, a dynamical energy of the space itself.

In \cite{grVort} the example of a globe having the
diameter 20 cm and rotating with 1 turn per second is studied. For draw the space around
it to coherent moving it needs to spend the energy of annihilation of 300 000
tons of matter.

\subsection{The Quantum  Gravity}

In a view of GTT, the quantum gravity theory undergoes the most significant modification.
The catastrophic relationship GR $H=0$, putting on hold all quantum dynamics, is removed.
In GTT, a quantum gravity, as well as quantum theory of other fields,
for example quantum electrodynamics, can be built on the basis of a
Schrodinger equation
\begin{equation} \label{Shred}
i\hbar \,\D{\Psi}{t} = \hat{H} \, \Psi,
\end{equation}
defining the dynamics {\it of a state vector} of space (and other fields) $\Psi$
in global time.

The measure of quantum fluctuations is thus defined not in some fixed space,
but by the metric of that curved space, in which these fluctuations arise.
Thus, in contrast for example to quantum electrodynamics,
where the basic problem turns out to be a nonlinearity (during interaction with a field of electrons)
but the functional space is flat, in quantum gravity the functional
space itself has a curvature \cite{Bur}.

Components of the metric $\gamma _{ij}$ commute with each other, as well as
components of momenta $\pi^{kl}$. However, an expression for the Hamiltonian
can be considerably simplified
 if one introduces {\it affine momenta} and subtracts from them a trace $ \pi^l_l $
 which commutes (in sense of Poisson brackets) with each affine momentum:
\begin{equation} \label{ssp}
 \pi^i_j=q^i_j+\frac{\delta^i_j}{3}\pi;\quad
q^i_i=0; \quad\pi^i_i =\pi,
\end{equation}
In these variables (and at $V^i=0$), the Hamiltonian (\ref{HamGr}) looks simpler:
\begin{equation} \label{afHam}
H=\frac{1}{\sqrt{\gamma}}\left(2\,q^i_jq^j_i-\frac{\pi^2}{3}\right)-
\frac{\sqrt{\gamma}}{2} \stackrel{(3)}{R},
\end{equation}
Note that the metric enters in the kinetic energy only through
$ \sqrt{\gamma} $, and this variable commutes with $q^i_j $, but the latter do not commute
 with each other:
\begin{equation} \label{qq}
\left\{q^i_j(x),q^k_l(x')\right\}=\frac{1}{2}(\delta^i_lq^k_j
\delta^k_j q^i_l) \delta (x-x ').
\end{equation}
This is the commutation relationship for currents of the $Sl(3)$ group, which thus naturally arises in
the dynamical theory of gravitation.

These complications arise as a sequence of the background independence of quantum gravity.

\subsection{Quantum cosmology}

As a trial stone in this or that approach to a quantum gravity
cosmological problems with a finite number of degrees of freedom are
often considered.

Let us consider a compact cosmological model of Friedman's type \cite{grQuant},
homogeneous and isotropic with space being a three-dimensional sphere, filled with
matter obeying the equation of state $ \varepsilon=3 \, p $.
As to the geometry of space, variations of the radius $r$ are only taken into account.

Hamiltonian
\begin{equation} \label{Hcosm}
 H=-\frac{p_r^2+r^2}{2\,r}+\frac{q^2}{2\,r},
\end{equation}
where $q^2 $ characterizes a conserved amount of a ultrarelativistic substance.

The wave function is a function of time and radius of a sphere $r $. The variables in this problem
can be separated. Denoting by a prime the derivative with respect to the radius
and simmetrizying $p^2/r $,
we obtain a stationary cosmological wave equation (in Planck units):
\begin{equation} \label{qeq}
u ''-\frac{u '}{r} + (-r^2+q^2) u=2 \, r \, E \, u.
\end{equation}

The question of a measure in a functional space requires an additional study (probably, $r^5 $),
but since we do this exercise just as a proof of principle, the equation (\ref{qeq})
is written for the measure in functional space $r^0=1$.

The spectrum of this equation is discrete. The solutions break up into two classes depending
on the behavior in a neighborhood of zero: as $r^0 $ and as $r^2 $.
The first class is of interest for study of a quantum problem of the Big Bang,
since in solutions of the second class probability density near zero radius is always equal to zero.

Eigenvalues of an energy for first eight such functions at $q=0$
(the matter is absent, only space dynamics is contributing) and $q=1 $ are listed in the table: \begin{center}
\begin{tabular}{|c|c|c |}
% after \\: \hline or \cline{col1-col2} \cline{col3-col4}...
\hline
\hline
n & q=0 & q=1 \\ \hline
1 & -0.977722 & 4.42817 \\
2 & -3.05247446 &-2.3182877 \\
3 & -4.16434141 &-3.6338011 \\
4 & -5.03491431 &-4.5990728 \\
5 & -5.77537028 &-5.3970940 \\
6 & -6.43100378 &-6.0924244 \\
7 & -7.02566164 &-6.7165684 \\
8 & -7.57373725 &-7.2876611 \\ \hline
\hline
\end{tabular}
\end{center}

At $q=0$, all eigenvalues of an energy are negative, at $q\geq 1 $ first modes
have a positive energy.

First six (unnormalized) functions for pure space ($q=0 $) are presented on the
graph:

\includegraphics [scale=0.7]{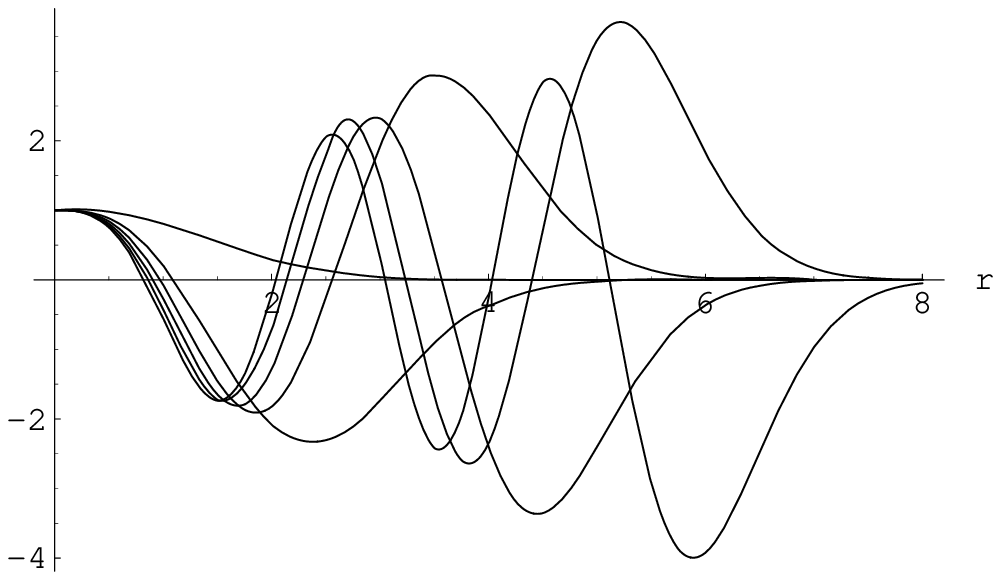}

Since the functions are not normalized and slightly notorthogonal due to
the approximate integration the metric, matrix is calculated
 $$M_{ij}=\int\limits_0^{r_{max}}u_i(r)\,u_j(r)\,dr$$
and the evaluation of matrix operators will be carried out with an inverse matrix
$K ^{ij} =M ^{-1} _{ij}$. The operator of the radius
$$ r^i_j =\sum _{s=1} ^n K^{is}\,\int\limits_0^{r_{max}}r\,u_s(r)\,u_j(r)\,dr.$$

is of interest for the analysis of dynamics of the radius. At $n=8$ eigenvalues
of this matrix are equal

( 0.51, 1.7, 2.6, 3.4, 4.2, 5.0, 5.8, 6.8).

With increase of $n$ (number of functions) the minimum eigenvalue decreases,
but the maximum one grows. The product of the two remains approximately slightly larger than $\pi$.
Therefore quantum effects do not prevent the Big Bang,
but accounting for the fact that the Plank's constant in Hevyside units has dimensionality
of the square of a length, lead to a hypothesis about some {\it a cosmological
uncertainty relation}:
\begin{quote}
The product of maximum and minimum radiuses of the World is not less than
$ \pi \,\hbar $.
\end{quote}

The wavelet in the space with $n=8 $, having a minimum eigenvalue
of radius $r=0.51 $ is presented in the figure.

\includegraphics [scale=0.7]{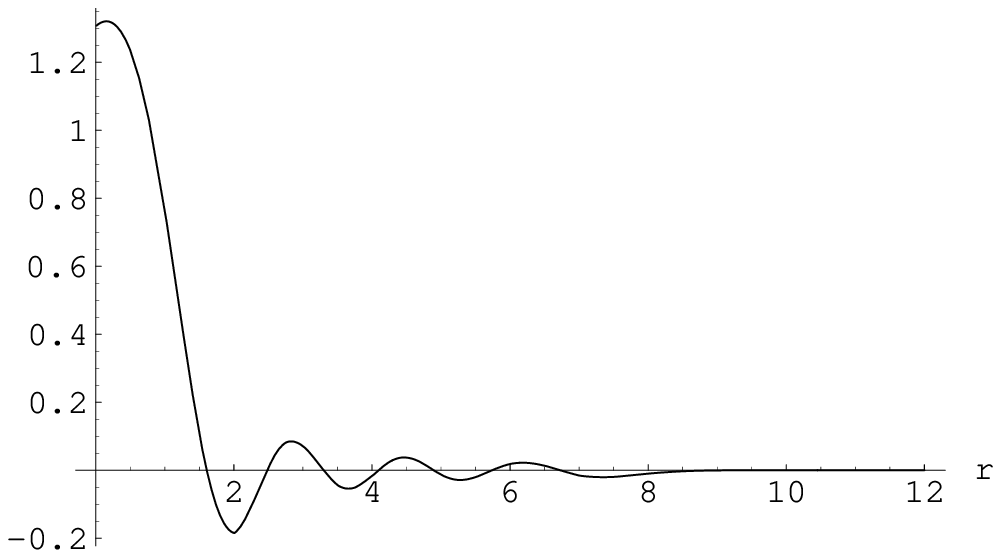}

However this wavelet is not stationary and begins to (widen) spread. This  expansion
is not monotonic: the wavelet (square of the module) breaks into two component,
one of which is moving away from zero, but second oscillates near to zero:

\vspace{8pt}
\includegraphics [scale=0.5]{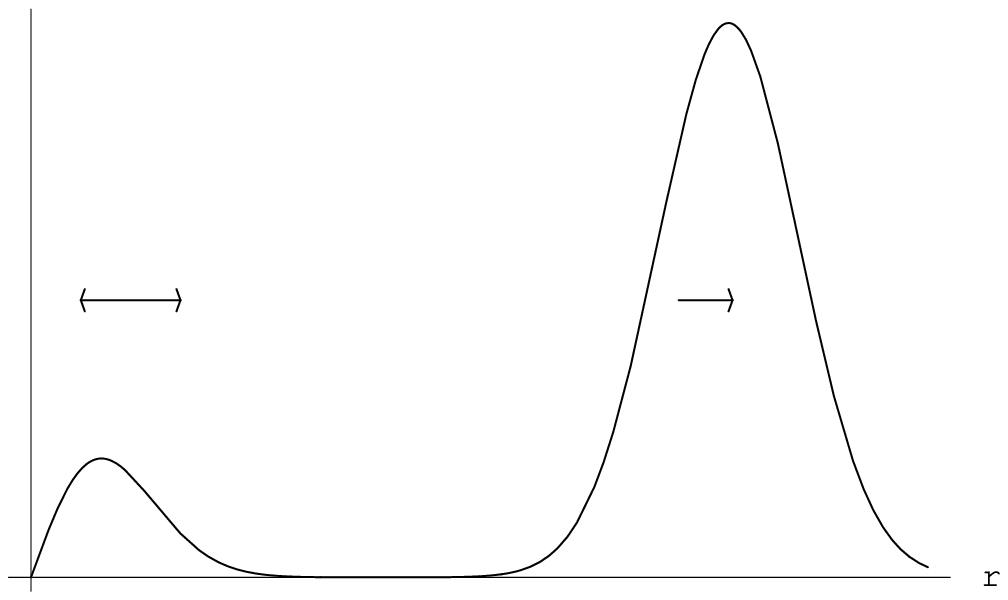}

From the point of view of a quantum mechanics in any problem (for instance solid state)
 this kind of behavior of the probability density would not cause serious
problems. However with respect to the Radius of the World (the unique, single variable) the problems
arise. What is the Radius of the World {\it for us?} While it is possible to reconcile to
average values and small fluctuations around, how then are we supposed to treat the simultaneous
probability of two essentially different radiuses?
What should we expect to get if we measure somehow the Radius of the World?
Will a reduction of a wavelet in the area of big,
or in the area of small radiuses happen?

The possible answer is, that {\it the Radius of the World cannot be measured directly}.
Hubble has determined it by measuring properties of photons,
coming from remote galaxies.
These photons also obey quantum theory and their quantum behavior
(accessible for our measurements)
can be influenced by both, the area of big and small radiuses,
but no reduction of a wave function of the Radius of the World will happen at fixing
a photon (see \cite{Blum}).

Overwhelming majority of quantum variables is being observed by nobody. Their quantum mechanical
behavior is exhibited only through their influences on a small number of observable
variables. From the quantum mechanical point of view, the Radius of the World  can have
quite {\it spread} values, but it can cause only specificity in observed
(or not observed yet) behavior of observed variables (photons, space particles).
Quite possible that quantum physics has in its basis not a wave function
but a density matrix.

\section{Conclusions}

Being built on more advanced mathematical framework,
GTT introduces into physics a new (in general old, known at a level
of philosophy for a long time) {\it physical} object: {\bf the space}. From the point of view
of theoretical physics, it is a nine-component field with curved functional space.
As well as other fields, for example electromagnetic, it has a density and flux
of energy, at the cosmic scale the energy of deformed space is enormous.

The study of properties of this physical object, will probably shed light
on modern problems of Cosmic dynamics which are currently being explained by "a dark matter"
and similar exotic substances. The study of quantum properties of space
will certain advance us in understanding a quantum essence of the World.

Physics of space is not interlinked tightely with the relativism
and can be under development to some extend {\it before} the special relativity theory.
This journey can be made following Niels Bj\"orn \cite{Bjern}, who although is
a fiction person, however the work with his articles
brought the author to the clear understanding of space dynamics in global time.
It is namely him, whom I express my deepest gratitude.

\vspace{12pt} %\bf

\end{document}